\documentclass{article}
\usepackage{epsfig}
\usepackage{amsmath}
\usepackage{color}
\newtheorem{theorem}{Theorem}[section]
\newtheorem{lemma}{Lemma}[section]

\newcommand{\blackslug}{\penalty 1000\hbox{
    \vrule height 8pt width .4pt\hskip -.4pt
    \vbox{\hrule width 8pt height .4pt\vskip -.4pt
          \vskip 8pt
      \vskip -.4pt\hrule width 8pt height .4pt}
    \hskip -3.9pt
    \vrule height 8pt width .4pt}}

\newenvironment{proof}{\vspace{1mm} \noindent {\sc Proof.}$\;$\rm}{\qed}
\newcommand{\qed}{\hspace*{\fill}\blackslug}
\def\boxit#1{\vbox{\hrule\hbox{\vrule\kern4pt
 \vbox{\kern1pt#1\kern1pt}
\kern2pt\vrule}\hrule}}
\begin{document}

\title{\bf On the Dominating Set Problem in Random Graphs}

\author{Yinglei Song \\
School of Electronics and Information Science\\
Jiangsu University of Science and Technology\\
Zhenjiang, Jiangsu 212003, China\\
syinglei2013@163.com\\
}
\date{}
\maketitle

\begin{abstract}

\noindent In this paper, we study the {\sc Dominating Set} problem in random graphs. In a random graph, each pair of vertices are joined by an edge with a probability of $p$, where $p$ is a positive constant less than $1$. We show that, given a random graph in $n$ vertices, a minimum dominating set in the graph can be computed in expected $2^{O(\log_{2}^{2}{n})}$ time. For the parameterized dominating set problem, we show that it cannot be solved in expected $O(f(k)n^{c})$ time unless the minimum dominating set problem can be approximated within a ratio of $o(\log_{2}n)$ in expected polynomial time, where $f(k)$ is a function of the parameter $k$ and $c$ is a constant independent of $n$ and $k$. In addition, we show that the parameterized dominating set problem can be solved in expected $O(f(k)n^{c})$ time when the probability $p$ depends on $n$ and equals to $\frac{1}{g(n)}$, where $g(n)< n$ is a monotonously increasing function of $n$ and its value approaches infinity when $n$ approaches infinity.
\end{abstract}

{\bf Keywords:} dominating set, random graphs, expected computation time, parameterized computation

\section{Introduction}

Given a graph $G=(V,E)$, a {\it dominating set} is a vertex subset $D \subseteq V$ such that each vertex in $V-D$ is adjacent to at least one vertex in $D$. The goal of the {\sc Minimum Dominating Set} problem is to find a dominating set of the minimum size in a given graph. It is well known that the problem is NP-hard \cite{garey}. It is therefore highly unlikely to develop algorithms that can solve the problem in polynomial time.

Recently, a few exact algorithms have been developed to break the trivial $O^{*}(2^n)$ bound in the time complexity of the minimum dominating set problem. In \cite{fomin}, it is shown that the number of minimal dominating sets in a graph in $n$ vertices is at most $O(1.7159^{n})$ and all these sets can be enumerated in $O(1.7159^n)$ time.  In \cite{fomin2}, it is shown that a minimum dominating set can be computed in $O(1.5264^n)$ time and polynomial space, the upper bound of the problem can be improved to $O(1.5137^n)$ in exponential space. The upper bound of the time complexity of this problem is further improved to be $O(1.5048^n)$ in \cite{rooij}. Although a large amount of efforts have been made to improve the worst-case time complexity of the problem, it remains unknown whether it can be solved in subexponential time or not. In \cite{chen}, it is shown that it is very unlikely to solve a few NP-hard problems in subexponential time, including the minimum dominating set problem.

Parameterized computation studies the possibility to develop efficient solutions to solve intractable optimization problems when the parameters in these problems are small integers. An NP-hard problem is {\it fixed parameter tractable} if it is associated with a few parameters $p_1, p_2, \cdots, p_m$ and can be solved in time $O(l(p_1, p_2, \cdots, p_m)n^{c})$, where $n$ is the size of the problem, $l$ is a function that only depends on the parameters and $c$ is a constant independent of the parameters. A fixed parameter tractable problem can be efficiently solved in practice if the parameters in the problem are all small integers.

For example, the {\sc Vertex Cover} problem is a well known fixed parameter tractable problem. The goal of the problem is to decide whether a given graph $G=(V,E)$ contains a vertex cover of size $k$ or not. Recent work has shown that the problem can be solved in time $O(1.2852^{k}+k|V|)$ \cite{chen0}. On the other hand, some NP-hard problems do not have known efficient parameterized solutions and are thus believed to be parameterized intractable. The parameterized dominating set problem is one of these problems. It has been shown that the problem is W[2]-complete \cite{downey1, downey2}, which implies that all problems in class W[2] are fixed parameter tractable if the problem is fixed parameter tractable. This fact suggests that it is highly unlikely to develop efficient parameterized algorithms for the parameterized dominating set problem.
A comprehensive survey on parameterized computation theory and complexity classes is provided in \cite{downey}.

Since it is unlikely to develop algorithms that can efficiently solve NP-hard optimization problems, a question arises naturally on whether approximate solutions with guaranteed approximation ratios can be obtained for these problems in polynomial time. In \cite{johnson} it is shown that the {\sc Minimum Vertex Cover} problem can be approximated within a ratio of 2.0 in polynomial time. For the {\sc Minimum Dominating Set} problem, a well known fact is that a greedy algorithm can achieve an approximation ratio of $\ln{n}$ in polynomial time \cite{johnson}, where $n$ is the number of vertices in the graph.

However, research in computational complexity also shows that some NP-hard problems cannot be approximated within a certain ratio unless NP=P. For example, it is shown in \cite{dinur} that the {\sc Minimum Vertex Cover} problem cannot be approximated within a ratio of $1.677$ unless NP=P. In \cite{hastad}, it is shown that it is NP-hard to approximate the maximum independent set of a graph with a ratio of $n^{1-\epsilon}$ in polynomial time, where $n$ is the number of vertices in the graph and $\epsilon$ is any positive constant. For the {\sc Minimum Dominating Set} problem, it has been shown that the problem cannot be approximated within a ratio of $c\ln{n}$, where $n$ is the number of vertices in the graph and $c$ is some positive constant independent of $n$ \cite{raz}. A similar inapproximability result is also obtained for the problem when the underlying graph is chordal or near chordal graphs \cite{liu0, liu}.

The Erd\H{o}s-R\'{e}nyi random graph model \cite{erdos} has provided valuable insights into many problems in both science and engineering. For example, recent research in molecular biology shows that the protein side-chain interaction network can be described by a random graph generated with the Erdos-Renyi model with high accuracy \cite{kuchaiev}. In this model, an edge appears between each pair of vertices independently with a probability of $p$, where $p$ is positive constant less than $1$. When $p=\frac{1}{2}$, each graph in $n$ vertices is generated with equal probability and the model thus provides an excellent platform to study the expected computational complexity of many NP-hard problems that are formulated on graphs.

In \cite{song}, we show that the {\sc Maximum Independent Set} problem can be solved in expected $2^{O(\log_{2}^{2}{n})}$ time in a random graph generated with the Erd\H{o}s-R\'{e}nyi model in $n$ vertices. In addition, it is shown that the parameterized independent set problem can be solved in expected time $2^{O(k^2)}+O(n^3)$ and the {\sc Maximum Independent Set} problem can be approximated within a ratio of $\frac{2n}{2^{\sqrt{\log_{2}{n}}}}$ in expected polynomial time. All these results suggest that the expected computational complexity of some NP-hard problems can be significantly different from its worst-case computational complexity. A question therefore arises immediately on whether similar results can be obtained for other NP-hard optimization problems or not, such as the {\sc Minimum Dominating Set} problem.

In \cite{nikoletseas}, an algorithm is developed to construct a near optimal dominating set for a random graph $G(n, p)$, where $n$ is the number of vertices in the graph and $p$ is a constant between $0$ and $1$ in expected polynomial time. In \cite{dryer}, it is shown that a minimum dominating set contains $O(\log_{2}{n})$ vertices for almost all random graphs generated with the Erd\H{o}s-R\'{e}nyi model in $n$ vertices. In \cite{wieland}, it is shown that the domination number of a random graph enjoys a concentration as sharp as its chromatic number. In \cite{glebov}, it is shown that the domination number of a random graph $G(n, p)$ is equal to one of the two values asymptotically almost surely whenever $p$ is a constant between $0$ and $1$. These results all reveal that the size of a minimum dominating set is $O(\log_{2}{n})$ in a random graph with a probability at least $1-\frac{1}{n^{O(1)}}$. However, these available lower bounds in probability are
not sufficient to lead to an exact algorithm that can compute a minimum dominating set in a random graph in expected subexponential time.

In this paper, we study the {\sc Dominating Set} problem in random graphs generated based on the Erd\H{o}s-R\'{e}nyi model. Using a new technique based on graph partition, we show that the problem can be solved in expected $2^{O(\log_{2}^{2}{n})}$ time. Although we are unable to develop an algorithm that can solve the parameterized dominating set problem in expected $O(f(k)n^c)$ time, we show that the expected parameterized complexity of the problem is inherently related to its approximability in expected polynomial time. In other words, we show that the parameterized dominating set problem can be solved in expected time $O(f(k)n^{c})$ if and only if the {\sc Minimum Dominating Set} problem can be approximated within a ratio of $o(\log_{2}{n})$ in expected polynomial time. In addition, our results also suggest that the problem is fixed parameter tractable in expected sense when the probability $p$ is a function of $n$ and equals to $\frac{1}{g(n)}$, where $g(n)<n$ is a monotonously increasing function of $n$ and its value approaches infinity when $n$ approaches infinity.

\section{Preliminaries and Notation}
Given a graph $G=(V,E)$ and vertex subset $D \subseteq V$,  a vertex $v \in G$ is {\it dominated} by $D$ if $v$ is adjacent to at least one vertex in $D$. The {\it domination number} of $G$ is the cardinality of a minimum dominating set in $G$. Let $m$ be a positive integer, a set of vertex subsets $P_1, P_2, P_3, \cdots, P_m$ in $V$ is {\it a disjoint $m$-partition} of $G$ if all $P_i$'s are pairwise disjoint and $\bigcup_{i=1}^{m}P_{i}=V$.

A partition $P_1, P_2, P_3, \cdots, P_m$ in $V$ is a {\it intersecting $m$-partition} of $G$ if $\bigcup_{i=1}^{m}P_{i}=V$ and it is not a disjoint $m$-partition of $G$. Given a vertex $v \in V$, the {\it degree} of $v$ in $G$ is the number of vertices that are adjacent to $v$ in $G$. Let $H=(U,F)$ be a directed graph, the {\it out-degree} of a vertex $u \in U$ is the number of directed edges that start with $u$ in $F$. The {\it in-degree} of $u$ is the number of directed edges that end with $u$ in $F$.

We use $G(n, p)$ ($0<p<1$) to denote a random graph generated with the Erd\H{o}s-R\'{e}nyi model in $n$ vertices. Each pair of vertices in $G(n, p)$ are joined with an edge independently with a probability of $p$.

\section{The Algorithm}

We start this section by showing that, with a probability of at least $1-2^{-O(n)}$, $G(n,p)$ contains a dominating set of size at most $O(\log_{q}n)$, where $q=\frac{1}{1-p}$. The proof uses a technique based on a disjoint partition of $G(n, p)$. We then refine the result with a more sophisticated technique using intersecting partitions of $G(n, p)$.

\begin{lemma}
\label{lm1}
\rm
Given a random graph $G(n,p)$ and a constant $C>1$, where $p$ is a constant between $0$ and $1$, with a probability at least $1-n^{2}2^{-\frac{C-1}{C}n\log_{2}{q}}$, $G(n,p)$ contains a dominating set of size at most $C\log_{q}{n}$, where $q=\frac{1}{1-p}$.

\begin{proof}
We assume that each dominating set in $G(n,p)$ is larger than $C\log_{q}n$. We then arbitrarily partition the vertices in $G(n, p)$ into a $l=\lfloor \frac{n}{C\log_{q}{n}} \rfloor +1$ disjoint partition $P_1, P_2, P_3, \cdots, P_{l}$ such that each of $P_1, P_2, \cdots, P_{l-1}$ contains $C\log_{q}{n}$ vertices and $P_{l}$ contains at most $C\log_{q}{n}$ vertices.

Given two subsets $P_i$ and $P_j$ in the partition, where $1 \leq i \leq l$, $1 \leq j \leq l$ and $i \neq j$, $P_i$ is {\it distinguished} by $P_j$ if $P_j$ contains a vertex $u$ such that $u$ is not dominated by $P_i$. Since each dominating set in $G(n, p)$ contains at least $C\log_{2}{n}+1$ vertices, each subset in the partition is distinguished by at least one other subset in the partition.

We now describe an algorithm that can be used to construct a directed graph $H$ from the subsets in the partition. Specifically, each subset in the partition is represented by a vertex in $H$ and $H$ thus contains $l$ vertices in total. We use $u_1, u_2, \cdots, u_l$ to denote the vertices that represent $P_1, P_2, \cdots, P_l$ respectively. $u_i$ is {\it distinguished} by $u_j$ if $P_i$ is distinguished by $P_j$. The steps of the algorithm can be described as follows.
\begin{enumerate}
\item{Set $M=\{u_1\}$ and $s=u_1$;}
\item{use $D_s$ to denote all vertices that distinguish $s$ and arbitrarily select a vertex $t \in D_s$ and create a directed edge from $s$ to $t$;}
\item{if $t$ is not in $M$, update $M$ to be $M \cup \{t\}$, set $s=t$ and go to step 2;}
\item{if $t \in M$, check whether $M$ contains all $l$ vertices in $H$ or not; if it is not the case, arbitrarily select a vertex $g$ that is not in $M$, set $s=g$ and go to step 2; otherwise, output $H$.}
\end{enumerate}
Since each vertex in $H$ is distinguished by at least one other vertex in $H$, the above algorithm constructs a directed graph $H$ such that the out-degree of each vertex is $1$. $H$ thus contains $l$ edges in total.

We now consider the probability associated with an edge in $H$. We assume the edge is from $u_i$ to $u_j$. We use $P(i, j)$ to denote this probability. It is clear that equation (\ref{eq1}) holds for $P(i, j)$.
\begin{equation}
\label{eq1}
P(i, j)=|P_j|(1-p)^{|P_i|}
\end{equation}
We then use $P(i)$ to denote the probability that $u_i$ is distinguished by a vertex in $H$. It is not difficult to see that the following inequality holds for $P(i)$.
\begin{eqnarray}
P(i) & \leq & \sum_{j=1}^{l}P(i, j) \\
     &   =  & \sum_{j=1}^{l}|P_j|(1-p)^{|P_i|} \\
     &   =  & n(1-p)^{|P_i|}
\end{eqnarray}
where the second equality is due to the fact that $\sum_{j=1}^{l}|P_j|=n$.

We use $E_i$ to denote the event that $u_i$ is distinguished by  another vertex as selected by the above algorithm in $H$. It is clear that $E_1, E_2, \cdots, E_l$ are mutually independent. We use $P$ to denote the probability that $G$ does not contain a dominating set of size $C\log_{q}n$. The following inequality holds for $P$.
\begin{eqnarray}
P & \leq & \prod_{i=1}^{l} P(i)  \\
  & \leq & n^{l}\prod_{i=1}^{l}(1-p)^{|P_i|} \\
  &  \leq & n^{\frac{n}{C\log_{q}{n}}+2} (1-p)^{n} \\
  &  = & n^{2} 2^{-\frac{C-1}{C}n\log_{2}{q}}
\end{eqnarray}
where the third inequality is due to the fact that $\sum_{i=1}^{l}|P_i|=n$ and $l \leq \frac{n}{C\log_{2}{n}}+2$. The lemma thus follows.
\end{proof}
\end{lemma}

It is shown in \cite{rooij} that a minimum dominating set can be computed in time $O(1.5048^n)$. Lemma \ref{lm1} thus suggests that when $p\geq \frac{1}{2}$, a large enough constant $C$ can be selected such that, with a probability of at least $1-o(1.5048^n)$, $G(n, p)$ contains a dominating set of size at most $C\log_{q}{n}$. A minimum dominating set in $G(n, p)$ can be computed with the following simple algorithm.
\begin{enumerate}
\item{Exhaustively enumerate all subsets in $G(n, p)$ that are of cardinality at most $C\log_{q}{n}$; for each such subset, check whether it is a dominating set of $G(n, p)$ or not;}
\item{if at least one dominating set is found, output the one that has the minimum cardinality; otherwise use the algorithm developed in \cite{rooij} to compute a minimum dominating set in $G(n, p)$ and output the result.}
\end{enumerate}
It is straightforward to see that the above algorithm can correctly compute a minimum dominating set in $G(n, p)$ in expected time $2^{O(\log_{2}^{2}{n})}$ when $p \geq \frac{1}{2}$.

However, it is also clear that Lemma \ref{lm1} is not sufficient to guarantee that the expected computation time of the above algorithm is $2^{O(\log_{2}^{2}{n})}$ when $p$ is a small constant. An improved lower bound for $P$ is thus needed to extend the above result to any constant $p$ that satisfies $0 < p < 1$.

We extend the disjoint partition we have used in the proof of Lemma \ref{lm1} to a series of intersecting partitions in $G(n, p)$. Based on these intersecting partitions, we show that the lower bound for $P$ can be significantly improved.

\begin{theorem}
\label{th1}
\rm
Given a random graph $G(n, p)$ and a constant $C>3$, where $p$ is a constant between $0$ and $1$, with a probability at least $1-n^{\frac{C}{2}\log_{q}{n}}2^{-(\frac{1}{6}-\frac{1}{2C})Cn\log_{2}{n}}$, $G(n, p)$ contains a dominating set of size at most $\frac{3}{2}C\log_{q}{n}$, where $q=\frac{1}{1-p}$.

\begin{proof}
We assume that any dominating set in $G(n, p)$ contains more than $\frac{3}{2}C\log_{q}{n}$ vertices. Similar to the proof of Lemma \ref{lm1}, a $l$ disjoint partition $P_1, P_2, \cdots, P_l$ can be obtained in $G(n, p)$, where $l=\lfloor \frac{n}{C\log_{q}n} \rfloor +1$, each of $P_1, P_2, \cdots, P_{l-1}$ contains $C\log_{q}{n}$ vertices and $P_{l}$ contains at most $C\log_{q}{n}$ vertices. We use $L_0$ to denote this partition.

Since any dominating set in $G(n, p)$ contains more than $\frac{3}{2}C\log_{q}{n}$ vertices, each subset $P_i$ in $L_0$ is distinguished by at least one other subset in $L_0$. We consider using the following algorithm to construct an intersecting partition $L_{1}$ in $G(n, p)$ from $L_{0}$.
\begin{enumerate}
\item{Color each vertex in every subset in $L_0$ to be red;}
\item{construct a directed graph $H$ as described in the proof of Lemma \ref{lm1};}
\item{set $M=\phi$, $S=\{u_1, u_2, \cdots, u_l\}$;}
\item{select a vertex $u_i \in S$, find the directed edge that starts with $u_i$ in $H$, we use $u_j$ to denote the other end of this directed edge;}
\item{since $u_i$ is distinguished by $u_j$, there exists a vertex $u \in P_j$ such that $u$ is not dominated by $P_j$. We update $P_i$ to be $P_i \cup \{u\}$ and color $u$ to be yellow in $P_i$ and change the color of $u$ to be green in $P_j$;}
\item{set $M=M \cup \{u_i\}$ and $S=S-\{u_i\}$;}
\item{go to step 4 if $S$ is not empty, otherwise output the partition formed by $P_1, P_2, P_3, \cdots, P_l$ as $L_1$.}
\end{enumerate}

It is clear that the partition returned by the above algorithm is an intersecting partition. In addition, different colors are assigned to vertices in each partition to clearly describe their different origins or roles with respect to the partition. Specifically, each vertex $u$ in $L_0$ is initially colored to be red. However, a vertex $u$ that is later added from $P_j$ to $P_i$ is colored to be yellow and its color in $P_j$ is changed from red into green. We note here that since $L_1$ is an intersecting partition, the color of a vertex may change when a different subset that contains the vertex is considered.

Let $L_k=\{P_{1}^{k}, P_{2}^{k}, \cdots, P_{l}^{k}\}$ be an $l$-intersecting partition in $G$ and each subset in $L_k$ contains at most $\frac{3}{2}C\log_{q}{n}$ vertices. We use the following algorithm to construct a new intersecting partition $L_{k+1}$.
\begin{enumerate}
\item{Set $M=\epsilon$, $S=\{P_{1}^{k}, P_{2}^{k}, \cdots, P_{l}^{k}\}$;}
\item{select a subset $P_{i}^{k}$ in $S$, arbitrarily choose a subset $P_{j}^{k}$ that distinguishes $P_{i}^{k}$ in $L_{k-1}$;}
\item{find a vertex $v_{i}^{k} \in P_{j}^{k}$ such that $v_{i}^{k}$ is not dominated by $P_{i}^{k}$;}
\item{update $P_{i}^{k}$ to be $P_{i}^{k} \cup \{v_{i}^{k}\}$, color $v_{i}^{k}$ to be yellow in $P_{i}^{k}$ and color $v_{i}^{k}$ to be green in
$P_{j}^{k}$;}
\item{update $M$ to be $M \cup \{P_{i}^{k}\}$ and $S$ to be $S-\{P_{i}^{k}\}$;}
\item{go to step 2 if $S$ is not empty, otherwise output $P_{1}^{k}, P_{2}^{k}, \cdots, P_{l}^{k}$ as $L_{k+1}$.}
\end{enumerate}

Since each subset in $L_k$ contains at most $\frac{3}{2}C\log_{q}{n}$ vertices, step 2 of the above algorithm can always find a subset $P_{j}^{k}$ that distinguishes $P_{i}^{k}$. We use $R_{k}$ and $R_{k+1}$ to denote the numbers of red vertices contained in $L_{k}$ and $L_{k+1}$ respectively. The following lemma establishes the relationship between them.

\begin{lemma}
\label{lm2}
\rm
$R_{k+1} \geq R_{k}-l$

\begin{proof}
We construct a directed graph $H_{k}$ to describe the relationships among subsets in $L_{k}$. A subset $P_{i}^{k}$ is represented with a vertex $u_{i}$ in $H_{k}$. A directed edge $(u_i, u_j)$ is created from $u_i$ to $u_j$ if $P_{i}^{k}$ is distinguished by $P_{j}^{k}$ and $P_{j}^{k}$ is selected by the above algorithm. It is not difficult to see that $H_{k}$ contains $l$ edges in total. The lemma follows from the fact that at most one red vertex is colored into green when an edge is created in $H_k$.
\end{proof}
\end{lemma}

For $L_{k}$, a subset $P_{i}^{k}$ is {\it partially distinguished} by another subset $P_{j}^{k}$ if $P_{i}^{k}$ does not contain red vertices or there exists a vertex $t_{j}^{k} \in P_{j}^{k}$ such that $t_{j}^{k}$ is not dominated by the set of red vertices in $P_{i}^{k}$. We use $E_{k}$ to denote the event that each subset in $L_{k}$ is partially distinguished by another subset in $L_{k}$. The following lemma establishes the independence of $E_{k+1}$ from $E_1, E_2, \cdots, E_{k}$.
\begin{lemma}
\label{lm3}
\rm
$E_{k+1}$ is independent of $E_1, E_2, \cdots, E_{k}$.

\begin{proof}
We consider an arbitrarily selected subset $P_{i}^{k}$ in $L_{k}$. We use $y_{i}^{1}, y_{i}^{2}, y_{i}^{3}, \cdots, y_{i}^{k}$ to denote the vertices that are included into $P_{i}^{k}$ by the construction of $L_{1}, L_{2}, \cdots, L_{k}$ respectively. It is straightforward to see that the vertex $v_{i}^{k}$ selected by the above algorithm is different from any of $y_{i}^{1}, y_{i}^{2}, y_{i}^{3}, \cdots, y_{i}^{k}$ since $v_{i}^{k}$ is not contained in $P_{i}^{k}$. The lemma thus immediately follows.
\end{proof}
\end{lemma}

\begin{lemma}
\label{lm4}
\rm
We use $P(E_{k})$ to denote the probability of $E_{k+1}$, then
$P(E_{k}) \leq n(1-p)^{(\frac{1}{3}-\frac{1}{C})n}$ for $0 \leq k \leq \frac{C}{2}\log_{q}{n}$, where $q=\frac{1}{1-p}$.

\begin{proof}
We use $R_{i}^{k}$ to denote the number of red vertices in subset $P_{i}^{k}$ in $L_{k}$. By Lemma \ref{lm2}, we immediately obtain the following inequality.
\begin{eqnarray}
    \sum_{i=1}^{l}R_{i}^{k} & \geq & n-kl \\
                            & \geq & \frac{n}{3}
\end{eqnarray}
where the second inequality follows from the fact that $k \leq \frac{C}{2}\log_{q}{n}$ and $l \leq \frac{n}{C\log_{q}{n}}+1$.

It is straightforward to see that the following inequality holds for $P(E_{k})$ when $0 \leq k \leq \frac{C}{2}\log_{q}{n}$.
\begin{eqnarray}
P(E_{k}) & \leq & \prod_{i=1}^{l}n(1-p)^{R_{i}^{k}} \\
         & \leq & n^{l}(1-p)^{\frac{n}{3}} \\
         & \leq & n^{1+\frac{n}{C\log_{q}{n}}}(1-p)^{\frac{n}{3}} \\
         & \leq & n(1-p)^{(\frac{1}{3}-\frac{1}{C})n}
\end{eqnarray}
The lemma thus follows.
\end{proof}
\end{lemma}

Starting with $L_0$, we can construct an intersecting partition $L_1$ with the first algorithm described in the proof. From $L_1$, we can construct $h$ intersecting partitions $L_2, L_3, \cdots, L_{h}$ in $G(n, p)$ where $h=\frac{C}{2}\log_{q}{n}$ with the second algorithm. Since each subset in $L_{k}$ contains at most $C\log_{q}{n}+k \leq \frac{3C}{2}\log_{q}{n}$ vertices, $E_{1}, E_{2}, E_{3}, \cdots, E_{h}$ must all occur. We use $P$ to denote the probability that $G(n, p)$ does not contain a dominating set of size $\frac{3}{2}C\log_{q}{n}$. From Lemma \ref{lm3} and Lemma \ref{lm4}, we immediately obtain
\begin{eqnarray}
P & \leq & \prod_{k=1}^{h}{P(E_{k})} \\
  & \leq & \prod_{k=1}^{h}n(1-p)^{(\frac{1}{3}-\frac{1}{C})} \\
  & \leq &  n^{h}[(1-p)^{(\frac{1}{3}-\frac{1}{c})}]^{h} \\
  & = & n^{\frac{C}{2}\log_{q}{n}}2^{-(\frac{1}{6}-\frac{1}{2C})nC\log_{2}n}
\end{eqnarray}
The theorem thus follows.
\end{proof}
\end{theorem}
From Theorem \ref{th1}, we can immediately obtain the expected computation time needed to compute a minimum dominating set in a random graph $G(n, p)$.

\begin{theorem}
\label{th2}
\rm
Given a random graph $G(n, p)$ where $p$ is a constant between $0$ and $1$, a minimum dominating set in $G(n, p)$ can be computed in expected time $O(2^{7\log_{q}{n}\log_{2}{n}})$, where $q=\frac{1}{1-p}$.

\begin{proof}
When $n$ is sufficiently large, we can set the constant $C$ in Theorem \ref{th1} to be $4$ and use the following algorithm to compute a minimum dominating set in $G(n, p)$.
\begin{enumerate}
\item{Exhaustively enumerate all vertex subsets of cardinality at most $6\log_{q}{n}$ and check if one of them is a dominating set of $G(n, p)$;}
\item{if at least one dominating set is found during the exhaustive search, return the one that is of the minimum cardinality; otherwise, use the algorithm developed in \cite{rooij} to compute a minimum dominating set in $G(n, p)$ and return the result.}
\end{enumerate}
By Theorem \ref{th1}, the algorithm developed in \cite{rooij} is called with a probability at most $2^{-O(n\log_{2}{n})}$. The theorem thus follows.
\end{proof}
\end{theorem}

\section{Expected Parameterized Complexity and Approximability}

In this section, we study the relationship between the expected parameterized complexity of the {\sc Dominating Set} problem  and the expected approximability of the {\sc Minimum Dominating Set} problem in random graphs. We show that they are in fact inherently related to each other. To establish the connection between them, we first show that the {\sc Minimum Dominating Set} problem can be approximated within a ratio of $o(\log_{2}n)$ in expected polynomial time if the parameterized dominating set problem can be solved in expected $O(f(k)n^c)$ time, where $k$ is the size of the dominating set and $n$ is the number of vertices in the graph.

The following lemma is needed to prove the connection in this direction.
It is originally proved in \cite{song}. We include the proof here for completeness.

\begin{lemma}
\label{lm5}
\rm
\cite{song}
Let $G(n, p)$ be a random graph and $\epsilon$ be a sufficiently small positive number, where $p$ is a constant between $0$ and $1$. With a probability at least $1-2^{-\mu n^2}$, there exists a vertex $u$ such that its degree in $G$ is at least $(p-\epsilon)n$, where $\mu$ is a positive constant that only depends on $\epsilon$ and $p$.

\begin{proof}
We assume such a vertex does not exist in $G(n, p)$, the graph thus contains at most $\frac{(p-\epsilon)n^{2}}{2}$ edges. However, it is straightforward to see that the expected number of edges in $G(n, p)$ is $\frac{pn(n-1)}{2}$. We use $P$ to denote the probability that such a vertex does not exist, an  upper bound for $P$ can be obtained based on the Chernoff bound.
\begin{eqnarray}
P & \leq & \exp{(-\frac{pn(n-1)}{4}\frac{(n\epsilon-p)^{2}}{(p(n-1))^{2}})} \\
  & \leq & \exp{(-\frac{\epsilon^{2}n^{2}}{32p})} \\
  & = & 2^{-\frac{\epsilon^{2}n^{2}}{32p\ln{2}}}
\end{eqnarray}
where the second inequality holds for sufficiently large $n$. The lemma follows by letting $\mu=\frac{\epsilon^{2}}{32p\ln{2}}$.
\end{proof}
\end{lemma}

Given a random graph $G(n, p)$ and a sufficiently small positive $\epsilon$, a vertex $u$ in $G(n, p)$ is {\it good} if its degree in $G(n, p)$ is at least $(p-\epsilon)n$. The following lemma states that a dominating set of size $O(\log_{2}{n})$ can be computed with a probability at least $1-2^{-O(\log_{2}^{2}{n})}$ in polynomial time.

\begin{lemma}
\label{lm6}
\rm
Let $G(n, p)$ be a random graph and $D$ be a positive constant, where $p$ is a constant between $0$ and $1$. With a probability at least $1-n2^{-\mu D^{2}\log_{2}^{2}{n}}$, a dominating set of size at most $\log_{\frac{1}{1-p+\epsilon}}n+D\log_{2}{n}$ can be found in polynomial time, where $\mu$ is some positive constant that only depends on $p$ and $\epsilon$.

\begin{proof}
We use the following algorithm to compute a dominating set in $G(n, p)$.
\begin{enumerate}
\item{Set $G_{1}=G$ and $S=\phi$;}
\item{return an empty set if $G_{1}$ does not contain a good vertex; otherwise find a good vertex $u \in G_{1}$;}
\item{update $S$ to be $S \cup \{u\}$ and remove $u$ and its neighbors from $G_1$;}
\item{check whether the number of vertices in $G_1$ is less than $D\log_{2}{n}$ vertices or not; if it is the case, include all the vertices in $G_1$ into $S$ and return $S$; otherwise go to step 2.}
\end{enumerate}
It is clear that the above algorithm either returns a dominating set $S$ of $G(n, p)$ or an empty set in polynomial time. Since a good vertex in $G_1$ is included in $S$ and all its neighbors are removed from $G_1$ when steps 2 and 3 are executed, the size of $G_1$ is at most $(1-p+\epsilon)^{r}n$ after steps 2 and 3 are executed for $r$ times. In addition, the algorithm returns when $G_1$ contains less than $D\log_{2}{n}$ vertices. Steps 2 and 3 are executed for at most $\log_{1-p+\epsilon}{n}$ times. The dominating set returned by the algorithm thus contains at most $\log_{1-p+\epsilon}{n}+D\log_{2}{n}$ vertices.

From Lemma \ref{lm5}, the algorithm terminates and returns an empty set in step 2 with a probability at most $2^{-O(\mu |V(G_1)|^{2})}$, where $\mu$ is a positive constant that only depends on $p$ and $\epsilon$ and $|V(G_1)|$ is the number of vertices in $G_1$. Since $G_1$ contains at least $D\log_{2}{n}$ vertices throughout the execution of the algorithm. From union bound, we immediately conclude that the algorithm returns an empty set with a probability at most $n2^{-\mu D^{2}\log_{2}^{2}{n}}$.
\end{proof}
\end{lemma}

\begin{theorem}
\label{th2}
\rm
Let $G(n, p)$ be a random graph, where $p$ is a positive constant between $0$ and $1$. The minimum dominating set in $G(n, p)$ can be approximated within a ratio of $o(\log_{2}{n})$ if there exists an algorithm that can determine whether $G(n, p)$ contains a dominating set of size $k$ or not in expected time $O(f(k)n^{c})$, where $c$ is a constant and $f(k)$ is a monotonously increasing function that only depends on $k$.

\begin{proof}
Since $f(k)$ is a monotonously increasing function that only depends on $k$, there exists a monotonously increasing function $e(n)$ such that $f(e(n))$ is bounded from above by a polynomial of $n$ for sufficiently large $n$. In addition, $e(n)$ approaches infinity as $n$ approaches infinity.

We use $A$ to denote the parameterized algorithm that can determine whether $G(n, p)$ contains a dominating set of size $k$ or not in expected time $O(f(n)n^{c})$. In addition, we use $B$ to denote the algorithm developed in the proof of Lemma \ref{lm6}. Choose a sufficiently small positive number $\epsilon$ and a positive constant $D$ such that $D > \sqrt{\frac{5}{\mu \log_{2}{q}}}$, where $q=\frac{1}{1-p}$ and $\mu$ is as defined in Lemma \ref{lm5}. We use the following algorithm to approximate a minimum dominating set in $G(n, p)$.
\begin{enumerate}
\item{Set the parameter $k=e(n)$ and use $A$ to determine whether $G(n, p)$ contains a dominating set of size $k$ or not; if the answer is ``yes'', set the parameter to be each integer between $1$ and $e(n)$ and apply $A$ on $G(n, p)$ for each parameter value to determine the size of the minimum dominating set in $G(n, p)$ and return the corresponding dominating set returned by $A$; otherwise continue to execute step 2;}
\item{use $B$ to find a dominating set in $G(n, p)$; return the dominating set returned by $B$ if it does not return an empty set; otherwise continue to execute step 3;}
\item{exhaustively enumerate all subsets of cardinality at most $4\log_{q}{n}$ in $G(n, p)$ and check whether a dominating set of $G(n, p)$ can be found in them. If it is the case, return the dominating set of the minimum cardinality; otherwise continue to execute step 4;}
\item{use the algorithm developed in \cite{rooij} to compute a minimum dominating set in $G(n, p)$ and return the result.}
\end{enumerate}

It is straightforward to see that the above algorithm can approximate a minimum dominating set in $G(n, p)$ within a ratio of $o(\log_{2}{n})$. Indeed, an exact solution is found if the algorithm returns in step 1, 3, or 4. An approximate solution is returned in step 2. However, we know from Lemma \ref{lm6} that the dominating set returned in step 2 contains at most $O(\log_{2}{n})$ vertices. In addition, the domination number of any $G(n, p)$ processed by step 2 is larger than $e(n)$ since the algorithm does not return in step 1. The approximation ratio of the algorithm is thus at most $O(\frac{\log_{2}{n}}{e(n)})=o(\log_{2}{n})$.

We then consider the expected computation time needed by the algorithm. Step 1 needs expected polynomial time since $f(e(n))$ is bounded from above by a polynomial of $n$. By Lemma \ref{lm6}, step 2 needs polynomial time.  The amounts of computation time needed by steps 3 and 4 are $O(n^{4\log_{q}{n}})$ and $O(1.5048^{n})$ respectively.

We use $ET(n)$ to denote the expected computation time of the algorithm and $ET_{3, 4}(n)$ to denote the expected computation time needed by the steps 3 and 4 altogether in the algorithm. In addition, we use $P_3(n)$ and $P_4(n)$ to denote the probabilities that steps 3 and 4 are executed respectively. By Lemma \ref{lm6} and Theorem \ref{th1}, we have $P_3(n) \leq n2^{-\mu D^{2}\log_{2}^{2}{n}}$ and $P_4(n) \leq 2^{O(-n\log_{2}{n})}$. The following inequality thus holds for $ET_{3, 4}(n)$.
\begin{eqnarray}
ET_{3,4}(n) & \leq & L(P_3(n)n^{4\log_{q}{n}}+P_4(n)2^{n}) \\
            & \leq & L(2^{-(\mu D^{2}-\frac{4}{\log_{2}{q}})\log_{2}^{2}{n}}+2^{n}2^{-O(n\log_{2}{n})}) \\
            & \leq & LH
\end{eqnarray}
where $L$ and $H$ are some constants independent of $n$. The third inequality is due to the fact that $D > \sqrt{\frac{5}{\mu \log_{2}{q}}}$ and $2^{n-O(n\log_{2}{n})}$ is bounded from above by a constant when $n$ is sufficiently large. The algorithm thus needs expected polynomial time.
\end{proof}
\end{theorem}

We then prove that the parameterized dominating set problem can be solved in expected $O(f(k)n^{c})$ time if the minimum dominating set problem can be approximated within a ratio of $o(\log_{2}{n})$ in expected polynomial time. We need the following lemma to establish the connection in this direction.

\begin{lemma}
\label{lm7}
\rm
Let $G(n, p)$ be a random graph, where $n$ is a sufficiently large integer and $p$ is a constant between $0$ and $1$. With a probability at most $\exp{(-\frac{\sqrt{n}}{4})}$, $G(n, p)$ contains a dominating set of size $\frac{\log_{q}{n}}{s(n)}$, where $s(n)$ is a monotonously increasing function of $n$ and its value approaches infinity when $n$ approaches infinity.

\begin{proof}
We assume $G(n, p)$ contains such a dominating set $D$. We use $P(D)$ to denote the probability for $D$ to exist. Since each vertex in $G-D$ is adjacent to at least vertex in $D$ and $G-D$ contains at least $\frac{n}{2}$ vertices when $n$ is sufficiently large, $P(D)$ must satisfy the following inequality.
\begin{eqnarray}
P(D) & \leq  & {n \choose \frac{\log_{q}{n}}{s(n)}} (1-(1-p)^{\frac{\log_{q}{n}}{s(n)}})^{\frac{n}{2}} \\
     & \leq & n^{\frac{\log_{q}{n}}{s(n)}}\exp{(-\frac{n^{1-\frac{1}{s(n)}}}{4})} \\
     & \leq & \exp{(-\frac{\sqrt{n}}{4})}
\end{eqnarray}
where the second inequality is based on the fact that $1-x < \exp{(-\frac{x}{2})}$ when $x$ is between $0$ and $1$. The third inequality holds for sufficiently large $n$ since $s(n)$ approaches infinity as $n$ approaches infinity. The lemma thus follows.
\end{proof}
\end{lemma}

\begin{theorem}
\label{th3}
\rm
Let $G(n, p)$ be a random graph, where $n$ is sufficiently large and $p$ is a constant between $0$ and $1$. There exists an algorithm that can determine whether $G(n, p)$ contains a dominating set of size $k$ or not in expected time
$O(f(k)n^{c})$ if the minimum dominating set problem can be approximated within a ratio of $o(\log_{2}{n})$ in expected polynomial time, where $f(k)$ is a function that only depends on $k$ and $c$ is a constant independent of $k$ and $n$.

\begin{proof}
We denote the approximate algorithm by $A$ and the approximation ratio of $A$ is $\frac{\log_{q}{n}}{w(n)}$, where $q=\frac{1}{1-p}$ and $w(n)$ is a monotonously increasing function. The value of $w(n)$ approaches infinity when $n$ approaches infinity. We use the following algorithm to determine whether $G(n, p)$ contains a dominating set of size $k$ or not.
\begin{enumerate}
\item{If $k>\sqrt{w(n)}$, use the algorithm developed in \cite{rooij} to compute a minimum dominating set in $G(n, p)$; return ``yes'' and the minimum dominating set found if $k$ is greater than or equal to the size of this dominating set; otherwise return ``no'';}
\item{otherwise, apply $A$ on $G(n, p)$ to obtain an approximate
solution $S$ for a minimum dominating set in $G(n, p)$;}
\item{if $|S| \leq \frac{\log_{q}{n}}{\sqrt{w(n)}}$, exhaustively search all vertex subsets of cardinality $k$ and check whether a dominating set can be found or not; if a dominating set is found return ``yes'' and the dominating set; otherwise return ``no''.}
\item{return ``no'' if $|S| > \frac{\log_{q}{n}}{\sqrt{w(n)}}$;}
\end{enumerate}

It is clear from the description of the algorithm that when $k>\sqrt{w(n)}$, the algorithm correctly determines whether $G(n, p)$ contains a dominating set of size $k$ or not in step 1. When $k< \sqrt{w(n)}$ and the cardinality of $S$ is larger than $\frac{\log_{q}{n}}{\sqrt{w(n)}}$, the domination number of $G(n, p)$ is larger than $\sqrt{w(n)}$ since the approximation ratio of $A$ is $\frac{\log_{q}{n}}{w(n)}$. The algorithm can thus recognize this case and correctly returns ``no'' in step 4. When $k < \sqrt{w(n)}$ and $|S| \leq \frac{\log_{q}{n}}{\sqrt{w(n)}}$, the algorithm correctly determines whether $G(n, p)$ contains a dominating set of size $k$ by exhaustive search in step 3.

The computation time needed in step 1 is at most $O(f(k))$, where $f(k)$ is a function of $k$. since when $k \geq w(n)$, there exists a function $z(k)$ such that $n \leq z(k)$. Step 1 thus needs $O(f(k))$ time. Step 2 applies $A$ to $G(n, p)$ and thus needs expected polynomial time. By Lemma \ref{lm7}, the probability for the algorithm to execute step 3 is at most $\exp{(-\frac{\sqrt{n}}{4})}$ when $n$ is sufficiently large. We use $ET_3(n)$ to denote the expected computation time needed by step 3. $ET_3(n)$ satisfies the following inequality when $n$ is sufficiently large.
\begin{eqnarray}
ET_3(n) & \leq & Ln^{k}\exp{(-\frac{\sqrt{n}}{4})} \\
        & \leq & Ln^{\sqrt{w(n)}}\exp{(-\frac{\sqrt{n}}{4})} \\
        & \leq & LM
\end{eqnarray}
where $L$ and $M$ are constants independent of $n$ and $k$. The second inequality holds since $k \leq \sqrt{w(n)}$; the third inequality holds since it is implied that $w(n)<\log_{q}{n}$ when $n$ is sufficiently large. The theorem thus follows.
\end{proof}
\end{theorem}

\section{Extension to Sparse Random Graphs}

In this section, we consider the case where the probability for an edge to appear between two vertices in $G(n, p)$ is not a constant. We assume $p=\frac{1}{g(n)}$, where $g(n)<n$ is a monotonously increasing function of $n$ and its value approaches infinity when $n$ approaches infinity. We show that the parameterized dominating set problem can be solved in expected $O(f(k)n^{c})$ time in this case. The proof is based on the well known $\ln{n}$-approximate algorithm for the minimum dominating set problem.

\begin{lemma}
\label{lm8}
\rm
Let $G(n, p)$ be a random graph and $p=\frac{1}{g(n)}$, where $g(n)<n$ is a monotonously increasing function and approaches infinity when $n$ approaches infinity. With a probability at most $\exp{(-\frac{\sqrt{n}}{4})}$, $G(n, p)$ contains a dominating set of size $g^{\frac{1}{3}}(n)\ln{n}$.

\begin{proof}
Similar to the proof of Lemma \ref{lm7}, we assume $G(n, p)$ contains such a dominating set $D$. We use $P(D)$ to denote the probability for $D$ to exist. Since each vertex in $G-D$ is adjacent to at least vertex in $D$ and $G-D$ contains at least $\frac{n}{2}$ vertices when $n$ is sufficiently large, $P(D)$ must satisfy the following inequality.

\begin{eqnarray}
P(D) & \leq & {n \choose  g^{\frac{1}{3}}(n)\ln{n}}(1-(1-p)^{g^{\frac{1}{3}}(n)\ln{n}})^{\frac{n}{2}} \\
     & \leq & n^{g^{\frac{1}{3}}(n)\ln{n}}(1-n^{g^{\frac{1}{3}}(n)\ln{(1-p)}})^{\frac{n}{2}}\\
     & \leq & \exp{(n^{\frac{1}{3}}\ln^{2}{n})}(1-\frac{1}{n^{\frac{2}{g^{\frac{2}{3}}(n)}}})^{\frac{n}{2}} \\
     & \leq & \exp{(n^{\frac{1}{3}}\ln^{2}{n})}\exp{(-\frac{1}{4}n^{1-\frac{2}{g^{\frac{2}{3}}(n)}})}  \\
     & \leq & \exp{(-\frac{\sqrt{n}}{4})}
\end{eqnarray}
where the third inequality follows from the fact that $\ln{(1-p)}>-2p$ when $n$ is sufficiently large; the fourth inequality is due to the fact that $g(n)<n$ and $1-x<\exp{(-\frac{x}{2})}$ holds for $x$ between $0$ and $1$; the last inequality holds when $n$ is sufficiently large.
\end{proof}
\end{lemma}

\begin{theorem}
\label{th4}
\rm
Let $G(n, p)$ be a random graph and $p=\frac{1}{g(n)}$, where $g(n)<n$ is a monotonously increasing function and approaches infinity when $n$ approaches infinity. There exists an algorithm that can determine whether $G(n, p)$ contains a dominating set of size $k$ or not in $O(f(k)n^{c})$ time, where $f(k)$ is a function of $k$ and $c$ is a constant independent of $n$ and $k$.

\begin{proof}
We use an algorithm similar to one developed in the proof of Theorem \ref{th4}. Since it is well known that a minimum dominating set can be approximated within a ratio of $\ln{n}$ with a greedy algorithm in polynomial time \cite{johnson}. We denote this approximation algorithm with $B$ and we use the following algorithm to determine whether $G(n, p)$ contains a dominating set of size $k$ or not.
\begin{enumerate}
\item{If $k>g^{\frac{1}{3}}(n)$, use the algorithm developed in \cite{rooij} to compute a minimum dominating set in $G(n, p)$; return ``yes'' and the minimum dominating set found if $k$ is greater than or equal to the size of this dominating set; otherwise return ``no'';}
\item{otherwise, apply $B$ on $G(n, p)$ to obtain an approximate
solution $S$ for a minimum dominating set in $G(n, p)$;}
\item{if $|S| \leq g^{\frac{1}{3}}(n)\ln{n}$, exhaustively search all vertex subsets of cardinality $k$ and check whether a dominating set can be found or not; if a dominating set is found return ``yes'' and the dominating set; otherwise return ``no''.}
\item{return ``no'' if $|S| > g^{\frac{1}{3}}(n)\ln{n}$;}
\end{enumerate}

Similar to the proof of Theorem \ref{th3}, when $k>g^{\frac{1}{3}}(n)$, the algorithm correctly determines whether $G(n, p)$ contains a dominating set of size $k$ or not in step 1. The computation time needed in this step only depends on $k$. When $k< g^{\frac{1}{3}}(n)$ and the cardinality of $S$ is larger than $g^{\frac{1}{3}}(n)\ln{n}$, the domination number of $G(n, p)$ is larger than $g^{\frac{1}{3}}(n)$ since the approximation ratio of $B$ is $\ln{n}$. The algorithm can thus recognize this case and correctly returns ``no'' in step 4. When $k < g^{\frac{1}{3}}(n)$ and  $|S| \leq g^{\frac{1}{3}}(n)\ln{n}$, the algorithm correctly determines whether $G(n, p)$ contains a dominating set of size $k$ or not by exhaustive search in step 3.

Step 2 applies $B$ to $G(n, p)$ and thus needs polynomial time. By Lemma \ref{lm8}, the probability for the algorithm to execute step 3 is at most $\exp{(-\frac{\sqrt{n}}{4})}$ when $n$ is sufficiently large. We use $ET_3(n)$ to denote the expected computation time needed by step 3. It is straightforward to see that $ET_3(n)$ satisfies the following inequality when $n$ is sufficiently large.
\begin{eqnarray}
ET_3(n) & \leq & Ln^{k}\exp{(-\frac{\sqrt{n}}{4})} \\
        & \leq & Ln^{g^{\frac{1}{3}}(n)}\exp{(-\frac{\sqrt{n}}{4})} \\
        & \leq & LM
\end{eqnarray}
where $L$ and $M$ are constants independent of $n$ and $k$. The second inequality holds since $k \leq g^{\frac{1}{3}}(n)$; the third inequality holds since $g(n)<n$. The theorem thus follows.
\end{proof}
\end{theorem}

\section{Conclusions}

In this paper, we study the dominating set problem in random graphs. We show that a minimum dominating set can be computed in expected $2^{O(\log_{2}^{2}{n})}$ time when the underlying graph is a random graph $G(n, p)$ generated based on the Erd\H{o}s-R\'{e}nyi model and $p$ is a constant between $0$ and $1$. In addition, we establish a close relation between the expected parameterized complexity and approximability of the problem in expected sense. We show that the parameterized dominating set problem can be solved in expected $O(f(n)n^{c})$ time if and only if the minimum dominating set problem can be approximated within a ratio of $o(\log_{2}{n})$ in expected polynomial time. With a similar approach, we show that the parameterized dominating set problem is fixed parameter tractable in expected sense when the probability $p=\frac{1}{g(n)}$, where $g(n)<n$ is a monotonously increasing function of $n$ and its value approaches infinity when $n$ approaches infinity.

Unfortunately, we are unable to develop an algorithm that can efficiently solve the parameterized dominating set problem in expected sense. In \cite{song}, we show that the independent set problem is fixed parameter tractable in expected sense. However, a well known fact is that the parameterized independent set problem is W[1]-complete while the parameterized dominating set problem is W[2]-complete. This suggests that the latter is more difficult to solve and it is therefore not surprising that the expected parameterized complexity of this problem cannot be easily improved. Our future work will focus on the possibility to improve the expected parameterized complexity of this problem.


\begin{thebibliography}{25}
\bibliographystyle{plain}
\bibitem{chen}
J. Chen, X. Huang, I. A. Kanj, and G. Xia, ``Linear FPT Reductions and
Computational Lower Bounds'', {\it Proceedings of the Thirty-Sixth ACM Symposium on Theory of Computing (STOC 2004)},
212-221, 2004.
\bibitem{chen0}
J. Chen, I. A. Kanj, and G. Xia, ``Improved Parameterized Upper Bounds for Vertex Cover'', {\it Proceedings of the Thirty-First International Symposium on Mathematical Foundations of Computer Science(MFCS 2006)}, 238-249, 2006.
\bibitem{dinur}
I. Dinur and S. Safra, ``The Importance of Being Biased'', {\it Proceeding of the Thirty-Fourth ACM Symposium on
Theory of Computing (STOC 2002)}, 33-42, 2002.
\bibitem{downey}
R. G. Downey and M. R. Fellows, {\it Parameterized Complexity}, Springer-Verlag, 1998.
\bibitem{downey1}
R. G. Downey and M. R. Fellows, ``Fixed Parameter Tractibility and Completeness i: Basic Theory'',
{\it SIAM Journal of Computing}, 24:873-921, 1995.
\bibitem{downey2}
R. G. Downey and M. R. Fellows, ``Fixed Parameter Tractibility and Completeness ii: Completeness for W[1]'',
{\it Theoretical Computer Science A}, 141:109-131, 1995.
\bibitem{dryer}
P. Dryer, Ph.D. Dissertation, Department of Mathematics, Rutgers University, 2000.
\bibitem{erdos}
P. Erd\H{o}s and A. R\'{e}nyi, ``On Random Graphs'', {\it Publicationes Mathematicae}, 6: 290-297, 1959.
\bibitem{fomin}
F. V. Fomin, F. Grandoni, A. Pyatkin, and A. Stepanov, ``Combinatorial Bounds Via Measure and Conquer: Bounding Minimal Dominating Sets and Applications'', {\it ACM Transactions on Algorithms}, 5(1): 9: 1-17, 2008.
\bibitem{fomin2}
F. V. Fomin, F. Grandoni, and D. Kratsch, ``A Measure and Conquer Approach for the Analysis of Exact Algorithms'', {\it Journal of the ACM}, 56(5):25:1-32, 2009.
\bibitem{garey}
M. R. Garey and D. S. Johnson, {\it Computers and Intractibility}, W. H. Freeman and Co., San Francisco, California, 1979. A
guide to the theory of NP-completeness, A Series of Books in the Mathematical Sciences.
\bibitem{glebov}
R. Glebov, A. Liebenau, and T. Szabo, ``On the Concentration of the Domination Number of the Random Graph'', {\it arXiv: 1209.3115}.
\bibitem{hastad}
J. Hastad, ``Clique is hard to approximate within $n^{1-\epsilon}$'', {\it Proceedings of the Thirty-Seventh Annual Symposium on Foundations of
Computer Science (FOCS 1996)}, 627-636, 1996.
\bibitem{johnson}
D. S. Johnson, ``Approximate Algorithms for Combinatorial Problems'', {\it Journal of Computer and System Sciences}, 9, 256-278, 1974.
\bibitem{kuchaiev}
O. Kuchaiev, T. Milenkovi\'{c}, V. Memi\v{s}evi\'{c}, W. Hayes, and N. Pr\v{z}ulj, ``Topological Network
Alignment Uncovers Biological Function and Phylogeny'', {\it Journal of Royal Society Interface},7(50):1341-1354, 2010.
\bibitem{liu0}
C. Liu and Y. Song, ``Parameterized Dominating Set Problem in Chordal Graphs: Complexity and Lower Bound'', {\it Journal of Combinatorial Optimization}, 18(1): 87-97, 2009.
\bibitem{liu}
C. Liu and Y. Song, ``Parameterized Complexity and Inapproximability of Dominating Set Problem in Chordal and Near Chordal graphs'', {\it Journal of Combinatorial Optimization}, 22: 684-698, 2011.
\bibitem{nikoletseas}
S. Nikoletseas and P. Spirakis, ``Near Optimal Dominating Sets in Dense Random Graphs'',{\it Proceedings of the 19th Workshop on Graph-Theoretic Concepts in Computer Science}, 1-10, 1993.
\bibitem{raz}
R. Raz and S. Safra, ``A Sub-Constant Error-Probability Low Degree Test, and a Sub-Constant
Error-Probability PCP Characterization of NP'', {\it Proceedings of the Twenty-Ninth ACM Symposium on Theory of Computing (STOC 1997)},
475-484, 1997.
\bibitem{rooij}
J. M. M. van Rooji, J. Nederlof, and T. C. van Dijk, ``Inclusion/Exclusion Meets Measure and Conquer: Exact Algorithms for Computing Minimum Dominating Sets'', {\it Proceedings of the Seventeenth Annual Symposium on Algorithms}, 554-565, 2009.
\bibitem{song}
Y. Song, ``On The Independent Set Problem in Random Graphs'', {\it International Journal of Computer Mathematics}, 92(11): 2233-2242.
\bibitem{wieland}
B. Wieland and A. Godbole, ``On The Domination Number of A Random Graph'', {\it Electronic Journal of Combinatorics}, 8(1): 37-49, 2001.
\end{thebibliography}
\end{document}